# Implications of New JHK Photometry and a Deep Infrared Luminosity Function for the Galactic Bulge

Glenn P. Tiede, Jay A. Frogel[1], and D. M. Terndrup[2]

Department of Astronomy, The Ohio State University,

174 W. 18th Ave., Columbus, OH 43210

Electronic mail: {tiede,frogel,terndrup}@payne.mps.ohio–state.edu

## ABSTRACT

We present deep near-IR photometry for Galactic bulge stars in Baade's Window, $(l, b) = (1.0°, -3.9°)$, and another minor axis field at $(l, b) = (0°, -6°)$. We combine our data with previously published photometry and construct a luminosity function over the range $5.5 \leq K_0 \leq 16.5$, deeper than any previously published. The slope of this luminosity function and the magnitude of the tip of the first ascent giant branch are consistent with theoretical values derived from isochrones with appropriate age and metallicity.

We use the relationship between [Fe/H] and the giant branch slope derived from near-IR observations of metal rich globular clusters by Kuchinski *et al.* [AJ, 109, 1131 (1995)] to calculate the mean metallicity for several bulge fields along the minor axis. For Baade's Window we derive $\langle[\text{Fe/H}]\rangle = -0.28 \pm 0.16$, consistent with the recent estimate of McWilliam & Rich [ApJS, 91, 749 (1994)], but somewhat lower than previous estimates based on CO and TiO absorption bands and the $JHK$ colors of M giants by Frogel *et al.* [ApJ, 353, 494 (1990)]. Between $b = -3°$ and $-12°$ we find a gradient in $\langle[\text{Fe/H}]\rangle$ of $-0.06 \pm 0.03$ dex/degree or $-0.43 \pm 0.21$ dex/kpc for $R_0 = 8$ kpc, consistent with other independent derivations. We derive a helium abundance for Baade's Window with the $R$ and $R'$ methods and find that $Y = 0.27 \pm 0.03$ implying $\Delta Y/\Delta Z = 3.3 \pm 1.3$.

Next, we find that the bolometric corrections for bulge K giants $(V - K \leq 2)$ are in excellent agreement with empirical derivations based on observations of globular cluster and local field stars. However, for the redder M giants we find, as did Frogel & Whitford [ApJ, 320, 199 (1987)], that the bolometric corrections differ by several tenths of a magnitude from those derived for field giants and subsequently used in the Revised Yale Isochrones. This difference most likely arises from the stronger molecular blanketing in the $V$ and $I$ bands of the bulge giants relative to that in field stars. From the luminosity function for Baade's Window and our bolometric corrections we calculate that $(m - M)_0 = 14.43 \pm 0.35$ or $R_0 = 7.7 \pm 1.2$ kpc to the bulge based on the surface brightness fluctuation method of Tonry & Schneider [AJ, 96, 807 (1988)], with the distance to M31 and M32 used for calibration.

## 1. Introduction

After Whitford's (1978) demonstration that the integrated spectrum of starlight in Baade's Window quantitatively resembles that of the nuclear spectra of many E and S0 galaxies, several research groups have actively been investigating the metallicities and ages of bulge stars, with the goal of understanding the stellar population in external systems.

The most essential component of models of the integrated light of galaxies, and the one that is the least accessible observationally, is the luminosity function (LF). Because the bulge is relatively nearby, we can determine the LF to faint magnitudes at several wavelengths, and measure the contribution of different types of stars to the integrated light in the bulge (Frogel & Whitford 1987; Terndrup $et$ $al.$ 1990). There is not at present a good determination of the LF in the near-infrared ($JHK$). Davidge (1991) derived a LF from a relatively small field in Baade's Window, but his photometry was not complete at faint magnitudes. DePoy $et$ $al.$ (1993) obtained $K$-band images of a large area in Baade's Window and derived a luminosity function; their photometry, however, was obtained with relatively low spatial resolution, did not go very deep, and did not include colors which can separate foreground from bulge stars (e.g., Terndrup 1988).

Another key ingredient in integrated light studies is the distribution of stellar metallicities. Until recently, it seemed that the basic metallicity scale was firmly established in Baade's Window, and that the majority of stars there had metallicities above solar, like that inferred for the nuclei of E and S0 galaxies. The high average metallicity was suggested by the profusion of late M giants; the complete absence of N-type carbon stars (Nassau & Blanco 1958; Blanco $et$ $al.$ 1984; Blanco & Terndrup 1989), by the CMD morphology from broadband optical photometry (Terndrup 1988; Geisler & Friel 1992), by infrared photometry including measures of CO absorption (Frogel & Whitford 1987; Frogel $et$ $al.$ 1990), and by medium resolution spectra of K and M giants (Whitford & Rich 1983;

Rich 1988; Terndrup *et al.* 1990, 1991). While it is true that a satisfactory quantitative explanation is lacking for a number of the differences between bulge giants and field giants, particularly in near-IR colors, these differences are in the expected sense for a population of stars with a mean metallicity equal to or greater than solar (Frogel 1993).

This seemingly straightforward picture of the metallicities of bulge giants may be unraveling. McWilliam & Rich (1994) obtained high resolution spectroscopy of K giants in Baade's Window, and suggested that the Rich (1988) metallicity scale should be revised downwards by about 0.3 dex to $\langle[\text{Fe/H}]\rangle = -0.27 \pm 0.40$. This finding is completely inconsistent with other indicators of overall metallicity, but McWilliam & Rich also noted that the relative abundance of Ti was enhanced compared to Fe. They suggested that this was the source of the strong TiO bands in the bulge M stars even though the mean iron abundance was less than solar. Most recently, Houdashelt (1995) has constructed a detailed stellar synthesis model based on the observations contained in Frogel & Whitford (1987 hereafter FW87) and in Terndrup *et al.* (1990). He compares this model to new observations of the central regions of E galaxies in the Virgo Cluster and concludes that the mean metallicity of the bulge is most similar to the metallicity of low luminosity E galaxies.

This paper is based on new $JHK$ photometry of two fields in the bulge. The paper is organized as follows: Section 2 of the paper describes our observations and data reduction techniques. In §3 we discuss our tests for completeness of the K-band photometry and the validity of the error estimates given by the DoPHOT package (Schechter *et al.* 1993). In §4, we discuss the color-magnitude and color-color diagrams of our fields. In addition, we present a new estimate for the mean metallicity in several bulge fields and discuss the the metallicity gradient in the bulge. In §5 we derive a $K$ band luminosity function for the bulge and compare it with previous work. We also describe a derivation of the bulge helium abundance based on the LF and discuss the validity of using the $R$ and $R'$ methods on a metal rich population like the bulge. In §6, we apply the Tonry-Schneider (1988) surface brightness fluctuation method to our LF, and derive a distance estimate of $7.7 \pm 1.2$ kpc to the bulge. A brief summary of the paper is contained in §7.

## 2. Observations, Image Reduction, and Photometry

### 2.1. Observations and Image Reduction

Our observations consist of deep $JHK$ images of two areas in the Baade's Window field $(l, b) = (1.0°, -3.9°)$, and one area in another field at $(l, b) = (0°, -6°)$, which we will call BW6 (the "BW" stands for "bulge window"). The areas imaged in Baade's Window are located to the northwest and to the east of the globular cluster NGC6522; we designate them respectively as BW4a and BW4b. These fields were chosen to include a significant number of stars with previous single-channel infrared photometry from Frogel & Whitford (1987) and optical photometry from Terndrup (1988) and Terndrup & Walker (1994). Table 1 lists the coordinates of the center of each field.

We imaged the three bulge fields with the IRCAM (Persson *et al.* 1992) on the 2.5m DuPont telescope at Las Campanas Observatory during the 1991 and 1993 observing seasons. The array in 1991 was a $128 \times 128$ HgCdTe NICMOS 2 detector with a scale of 0.52 arcsec pixel$^{-1}$. In 1993, the detector was a $256 \times 256$ HgCdTe NICMOS 3 detector with a scale of 0.348 arcsec pixel$^{-1}$. Individual exposure times and the total number of coadded frames are given in Table 2. The observations reported in this paper for the BW4a and BW4b fields were all obtained with the NICMOS 3 detector, while the BW6 data were obtained with the NICMOS 2. We also obtained some BW4a and BW4b frames with the latter detector. Although these frames are generally inferior to the ones we discuss here, they were useful for calibration of the BW6 data. The photometry reaches to $K \approx 16.5$ in the BW4b and BW6 fields, and to $K \approx 14.5$ in the BW4a field.

Figures 1a–c display representative $K$ band images of BW4a, BW4b and BW6 respectively. Figures 1b and 1c are deep $K$ band images, presented to show the depth and degree of crowding in our data. Note the electronic "ghosts" (circled in white in Figure 1b) caused by saturated pixels on the NICMOS 3 chip. These regions and affected stars were removed by hand after photometric processing. Labeled stars are stars from Blanco *et al.* (1984) and are the stars used to calibrate the photometry (§2.2). Figure 1d is an artificial data frame created to mimic a deep BW4b $K$ frame (see §3.1).

We used the same procedures as described in Kuchinski *et al.* (1995) to prepare the $JHK$ images for photometry. Briefly, we first applied a correction for the slightly nonlinear response of the detectors. Dark frames of the appropriate exposure time were subtracted from all object and sky frames. A flat field frame was constructed by median filtering all sky frames that were taken interspersed with the object frames. Finally we aligned the several exposures in each field, then averaged them. Since the data are well sampled, (typically each image had stellar FWHM $\geq 3.0$ pixels), integer pixel shifts were found to be satisfactory for aligning the frames. All steps of this process were performed using routines in IRAF.

## 2.2. Photometry

We obtained instrumental magnitudes with DoPHOT (Schechter *et al.* 1993). Before processing, we found that it was necessary to mask out the cores of saturated stars to prevent confusion of DoPHOT's saturated star algorithm. After processing, stars located within 2 stellar FWHM of the frame edges were eliminated from subsequent analysis because they were found to have unreliable photometry. Due to the large number of individual frames in each coadded image (as great as 50), the PSFs did not vary significantly with position on the coadded frames.

We calibrated the BW4a and the short exposure BW4b frames using previously determined magnitudes and colors for 11 stars from single-channel photometry (FW87). To calibrate the deep BW4b frames, the calibration of the short exposure frames was transferred via a comparison of all unsaturated stars in common. One night of our observations (UT 1993 July 5) was photometric, which allowed us to set up additional secondary standards in these two fields. As discussed in Kuchinski *et al.* (1995), there is no color term required to put the instrumental IRCAM colors and magnitudes onto the CTIO/CIT system of Elias *et al.* (1982, 1983). Thus on any given night, magnitudes and colors on the CTIO/CIT system are equal to the observed values plus a constant. Since the constant was determined independently for each frame it includes any necessary airmass correction. We then calibrated the other BW4a and BW4b observations by determining the magnitude offsets to match the calibrated magnitudes for all stars in common.

For the BW6 field, we adopted a two-step process. We first applied standard airmass corrections to the instrumental magnitudes, then determined the aperture magnitude offsets between the BW6 stars and observations of the BW4 fields taken on the same observing run (in 1991; these frames are not otherwise discussed in this paper). We then determined the transformation between these NICMOS 2 frames in Baade's Window to the secondary standards determined as discussed above.

The last three columns of Table 2 list the uncertainty in the zero-points for each field, which were computed from the unweighted standard deviation in the determination of the zero point. The dominant source of error is image crowding. Tables 3, 4, and 5 present our final calibrated photometry for BW4a, BW4b, and BW6 respectively. The XY pixel scale is for the final combined $K$ photometry with an arbitrary zero point. Stars used as standards are indicated in the notes column with their Arp (1965) or Blanco *et al.* (1984) ID codes and are also indicated on Figures 1a and 1b.

## 3. Photometric Error Analysis and Completeness

In order to derive a luminosity function at faint magnitudes (see §5), we need to determine quantitatively the photometric accuracy and star detection efficiency of our photometric measurements. To do this a known number of artificial stars of known magnitude must be introduced to an image and then measured using a procedure identical to the procedure used with the real data. However, adding stars to our already crowded images significantly increases the crowding, the major limiting factor to both photometric accuracy and star detection. Therefore rather than simply adding stars to our existing data frames, we constructed entire artificial data frames carefully mimicking the real data frames, and then reduced them in a manner identical to the original data reduction.

### 3.1. Generation of Artificial Data

IRAF routines contained in **artdata** were used to construct images with the same area, sky level, gain, and read noise as a deep BW4 or BW6 data frame. The stellar density was simulated with the power law luminosity function derived by DePoy *et al.* (1993), and was normalized to reproduce the observed number of stars in the range $9 \leq K \leq 16.5$. This power law, with the same normalization, was extrapolated over the range $16.5 \leq K \leq 20$ as an estimate for the unresolved bulge background. These background stars were included in the artificial frame to correctly reproduce sky levels. Scaled and shifted empirical stellar profiles from the real frames were used for each artificial star. These stars, a sky level and a simulated readout noise were added along with photon noise appropriate for the sky and star values were added to form the simulated frames. A number of simulated frames equal to the number of real frames in the real final images were generated and then coadded, to simulate the real noise statistics for each artificial image.

Even after taking these steps to model the noise, a plot of flux per pixel versus number (i.e. a flux histogram) of the artificial frame did not match the flux histogram of the real frame in two ways. Both histograms resembled a Poisson distribution as expected and had identical modal values, but the width of the peak of the artificial data was narrower than the width of the of the peak of the real data. By examining the flux histogram of small regions of the frame, we determined that the extra width of the real data was due to the fact that the real frame was not perfectly flat. The sky level was varying about the modal value and so appeared as increased dispersion in the flux histogram. Secondly, we noted a slight deficit in the number of pixels in the high wing of the artificial data flux histogram relative the the real data histogram. This difference arose because simulated saturated stars were simply scaled stellar profiles and so did not correctly mimic the saturated stars in the real data. However, this difference was non-consequential for two reasons. First, less than 1% of the frame was affected. Second, those pixels that were affected were in the wings of saturated stars. DoPHOT deals with saturated stars by simply excising a square region around the star. Therefore, the affected pixels were deleted during the photometric measurements, and an equal number of pixels were similarly deleted in the simulated data.

After determining that the simulated frames were identical in all important ways to the real data, the simulated frames were run through DoPHOT with the same procedure used for the real data.

### 3.2. Photometric Accuracy and Error Analysis

One aspect of DoPHOT photometry which is important and has not yet been examined in the literature is the reliability of the photometric error assigned each star. DoPHOT assigns an error by scaling the $\sigma^2$ of the fit (given by the reciprocal of the diagonal term of the curvature matrix used in fitting the PSF to the star) by the reduced $\chi^2$ (Schechter *et* *al.* 1993). Therefore the reliability of the DoPHOT error estimate is a function of the signal to noise in the stellar profile, the similarity of the PSF being fit to the actual PSF of the stars, and the accuracy to which the noise in the frame is being modeled. Image crowding affects all of these.

Figure 2 and Table 6 display the difference between input and measured $K$ magnitudes as a function of input $K$ magnitude for the artificial frame. Errors bars show the rms difference in each 0.25 magnitude bin. On average the photometry is good to $\sim 0.03$ magnitudes and differences do not become systematic until $K > 16.25$ at which point DoPHOT measures a magnitude which is too bright. This finding is consistent with results in Schechter *et al.* (1993), and is caused by an underestimation of the background in crowded fields.

Figure 3 shows that the scatter in the true errors (input magnitude − measured magnitude) is much larger than the scatter in the errors assigned by DoPHOT, (Figures 3a and 3b respectively). However, the assigned error is just an estimate of the true error so to make a meaningful comparison the *average* of the true errors must be compared to the assigned errors. The histogram in Figure 3a is the average of the true errors in bins 0.25 magnitudes wide. This histogram is plotted over the assigned errors in Figure 3b for comparison. Overall, the histogram follows the trend of the assigned errors fairly well, with a possible deviation in the dimmest three bins which are below the completeness limit as shown in the next section. Therefore, the errors given by DoPHOT are a good approximation to the true errors, even in a field as crowded as ours.

### 3.3. Completeness

Figure 4a shows a histogram of the number of stars input into the artificial frame, and the number of these stars detected and measured by DoPHOT. The noise in the input

– 12 –power law is from counting statistics alone; IRAF randomly samples the input power law to determine the magnitudes of the artificial stars. Inspection shows that completeness begins to fall off in the $16 - 16.25$ magnitude bin. Examination also shows that few of the bins in the entire magnitude range are 100% complete (Table 6), and that the brightest bin is only 28% complete. All of these effects are explained below.

The brightest bin is incomplete because it contains the saturation limit. Of the 7 stars in the $12 - 12.25$ magnitude bin, only two were of low enough flux and positioned relative to the pixel grid such that the brightest pixel contained less than 25000 counts. Since we have simulated a deep frame, this saturation limit is not the saturation limit of our photometry overall, but rather the saturation limit of the deep frame only.

Few of the bins are 100% complete because DoPHOT handles saturated stars by deleting the affected pixels. Since the behavior of detectors at or near the saturation limit is somewhat unpredictable, rather than trying to fit and subtract a saturated star, DoPHOT simply excises a rectangular region around each such star. The size of this region was set to excise out to a distance where the wings of a saturated star were less than $2\sigma$ above the sky. In dealing with saturated stars in this manner, DoPHOT guarantees that they do not corrupt the photometry of other stars. Unfortunately, this process also deletes any stars that happen to fall near a saturated star. This is a non-systematic effect and as such does not select for or against stars in any way other than their proximity to a saturated star. However, this is the effect that causes few bins to be 100% complete. This effect is perhaps more apparent in Figure 4b which is a plot of the completeness in each 0.25 magnitude bin. Of the 15 bins between 12.25 and 16.00 magnitudes the average completeness is $93\% \pm 6\%$. This number is in excellent agreement with the percentage of the artificial image that was not excised by DoPHOT, 92.6%. Therefore, the photometry is considered complete to the $K = 16$ bin.

The first bin to depart significantly from the 93% level of completeness is the 16 - 16.25 magnitude bin. With a completeness of 77%, it is more than $2\sigma$ below the 93% level. The next bin is only 50% complete, so clearly the cut-off is being reached. Figure 5 shows that the real data behaves very similarly to the artificial data. No normalizations or corrections have been made to either histogram beyond the normalization of the number of stars input in the magnitude range $9.0 \leq K_0 \leq 16.5$, yet they agree very well in three important ways. In addition to the general trend, which shows that to within the counting statistical error the input power law was a good approximation, both the saturation limit and limiting magnitude fall-offs behave very similarly. Therefore the real data photometry is also considered complete to $K = 16$. Due to the steep fall off in completeness, 100% to 15% in one magnitude (see Table 6 as well as Figure 4b), no attempt is made to correct any bins at $K > 16$.

An identical process was carried out mimicking a long $K$ exposure of the BW6 field. The stellar density in BW6 is lower and the effective exposure times were similar (the BW4 and BW6 fields were taken with different detectors; see §2.1) so the completeness limit was expected to be comparable or somewhat deeper. We found that the photometry of the artificial image was complete to $\sim 16.25$ magnitudes. As in the BW4b case, comparison of the real data with the artificial data showed very similar behavior. Therefore, the completeness limit was confidently extended to the real data.

We are now in a position to derive and discuss the luminosity function for the bulge fields. This is done in §5 below.

## 4. Color-Magnitude and Color-Color Diagrams

### 4.1. Morphology of the Color-Magnitude Diagrams

We display our color-magnitude diagrams (CMDs) in $(J-K)_0$, $K_0$ in Figures 6 (BW4a and BW4b fields) and 7 (BW6). The open squares show single-channel photometry from FW87 and Frogel *et al.* (1990), the pluses display the single-channel photometry from Frogel *et al.* (1984), and the filled points are from this study. The points from the present study are weighted mean values, where the weights are determined from the exposure times, after eliminating the dim large-error stars from the short exposures. The colors and magnitudes have been corrected for reddening using the values adopted by FW87 and by Frogel *et al.* (1990) (see Table 1). The solid lines are linear fits to the giant branches and were used to determine the mean metallicity in each bulge field as discussed in § 4.2.

The CMDs of both BW4b and BW6 extend to $K_0 \approx 16.5$ magnitudes, about 3 magnitudes deeper than the study of DePoy *et al.* (1993) and about the same depth as that obtained by Davidge (1991), although the completeness limit for our data is $\sim 2.5$ magnitudes deeper than Davidge's. The data for BW4a reach only $K_0 \approx 14.5$ because the integration times were only $\sim 5\%$ as long as for the BW4b field.

In general the features on our infrared CMDs are identical to those seen in optical CMDs of the bulge: we detect the sequence of stars which has been attributed to blue foreground dwarfs (from $K_0 \sim 13$ to $K_0 \sim 16.5$ with $(J-K)_0 \leq 0.45$), an extended giant branch, and a giant-branch clump near $K_0 = 13$ (e.g., Terndrup 1988; Tyson 1991; Geisler & Friel 1992; Terndrup & Walker 1994; Paczynski *et al.* 1994; Stanek *et al.* 1994). The latter feature is particularly obvious when we plot the luminosity function (§5.1).

Although Davidge's (1991) IR photometry goes as deep as ours, he found no evidence for a foreground sequence, and assigned the relatively blue stars he found between $12 < K_0 < 14$ to the horizontal branch. We can demonstrate that this is incorrect by plotting in Figure 8 the CMD for Baade's Window in $I_0$, $(V-I)_0$. This CMD was constructed from the photometry of Terndrup & Walker (1994), which overlaps our BW4b field. We define the region of foreground dwarfs to be $I_0 \leq 15.75$, $(V - I)_0 \leq 0.7$ and designate it "region I" on the CMD; we then plot stars in that region with both optical and infrared photometry in Figure 9, a CMD in $K_0$, $(J - K)_0$, as filled squares. The distribution of these stars on *both* CMDs is along a nearly vertical line, quite unlike what is expected for a horizontal branch.

Similarly, we also note on Figures 8 and 9 that the stars in the giant-branch clump in the optical (Region II on Figure 8) are the same as those in the clump in the infrared (filled circles on Figure 9). Excluding the brightest three stars common to both clumps (one of which has very large errors in its optical magnitudes), the remaining stars in common to both extend from $K_0 = 12.6$ to 13.75 in the infrared CMD. The five stars that fall in this range in the infrared CMD that are not marked as corresponding with stars in the optical Region II were not detected optically. Finally note that because of their blueness, the field stars are generally fainter than the clump stars in Figure 9, whereas they are of comparable brightness in Figure 8.

## 4.2. The Giant Branch Slope and the Bulge Metallicity Gradient

Our infrared CMDs can be used to estimate the mean metallicity of the giants in the three observed fields, and in combination with the data from Frogel *et al.* (1990), the magnitude of the metallicity gradient along the minor axis of the bulge. We assume that the technique introduced by Kuchinski *et al.* (1995) is valid for bulge stars. They noted that the slopes of the upper giant branches of metal-rich globular clusters increased with metallicity, and derived

$$[\text{Fe/H}] = -2.98(\pm 0.70) - 23.84(\pm 6.83) \times \frac{\Delta(J-K)}{\Delta K} \qquad (4\text{-}1)$$

over the range $-1.01 \leq [Fe/H] \leq -0.25$. We determined the giant-branch slope using the combined array and single-channel data, over the range $8.0 \leq K_0 \leq 12.6$, excluding obvious outlying points. These fits are shown superimposed upon the CMDs in Figures 6 and 7. The limits for the fits are the same in absolute $K$ magnitude as those used by Kuchinski *et al.* (1995). The faint limit excludes HB and clump stars, while the bright limit excludes the curvature in the upper giant branch, which is most likely due to the AGB.

We also computed the giant-branch slope for other fields along the minor-axis of the bulge from the single-channel photometry in Frogel *et al.* (1990). These are shown in Figure 10. The single-channel photometry for the other bulge fields does not extend as deep as our new data, so for those fields we fit the slope using photometry down to $K_0 \sim 11$. This does not introduce any systematic error, however, since for the giant branches in those fields with deeper photometry, the slopes measured down to $K_0 = 11$ and to $K_0 = 12.6$ were always equal within the errors. Table 7 displays the values of the slopes and the calculated metallicity estimates for the bulge fields. We also include slopes and estimates of the mean metallicity of the bulge fields near the globular clusters Terzan 2 (Kuchinski *et al.* 1995) and Liller 1 (Frogel *et al.* 1995). The giant-branch slope for fields BW4a and BW4b are equal to well within the uncertainties in the fit, thus, only one value for BW4 is tabulated.

Shown in Table 8 are previous determinations for $\langle[Fe/H]\rangle$ in several bulge fields taken from the literature (Terndrup 1988; Frogel *et al.* 1990; Tyson 1991; and McWilliam & Rich 1994). We have applied some corrections to these values as described in the notes to Table 8. Our estimates of $\langle[Fe/H]\rangle$ are $-0.28 \pm 0.16$ for BW4 and $-0.65 \pm 0.18$ for BW6. The uncertainties in these $\langle[Fe/H]\rangle$ values are formal uncertainties from the fit propagated through the metallicity equation. While the mean metallicity in Baade's Window is somewhat lower than the previous estimate of Frogel (1988), it is in good agreement with the other studies, particularly with the recent determination based on high-resolution spectra (McWilliam & Rich 1994) of $\langle[Fe/H]\rangle = -0.27 \pm 0.40$.

In Figure 11, we plot the $\langle[\text{Fe/H}]\rangle$ along the minor axis of the bulge as a function of $|b|$ for the four previous studies from the literature and for our new [Fe/H] estimates. For the new estimates (eq. 4.1), Figure 11a displays a linear fit to $\langle[\text{Fe/H}]\rangle$ as a function of $|b|$, and has a slope of $\Delta\langle[\text{Fe/H}]\rangle/\Delta|b| = -0.060 \pm 0.033$ dex/degree ($-0.43 \pm 0.21$ dex/kpc for $R_0 = 8$ kpc), in good agreement with the average slope found in previous investigations. The slopes from the previous investigations also generally agree with the least squares fit to the combination of all of the data from the four previous studies plus our new data (dotted lines in Figure 11), as is shown in Table 6. This agreement among the determined metallicity gradients and individual metallicity values, is consistent with our assumption that the Kuchinski *et al.* (1995) technique is measuring the same mean metallicity in the bulge that it measures in globular clusters.

### 4.3. Color-Color Diagrams

In Figure 12 we plot a $(J-H)_0$, $(H-K)_0$ diagram of BW4. The open circles are data from this survey, while the $\times$ symbols are single-channel photometry of Baade's Window M giants (FW87). The solid line is the mean locus for solar neighborhood giants and the dotted line is the locus for solar neighborhood dwarfs (Bessell & Brett 1988). The filled points indicate stars identified as lying in the foreground disk sequence (from Figures 8 and 9).

The distribution of stars with $(J-H)_0 \geq 0.4$ on this diagram is consistent with that found in previous studies of the bulge (e.g., FW87). In particular, the M giants (stars with $(J-H)_0 \geq 0.6$) lie to the red of the solar-neighborhood sequence in $(H-K)_0$ and to the blue in $(J-H)_0$, which seems to continue the trend with increasing metallicity from globular cluster giants to giants in the solar neighborhood (FW87; Frogel 1988; Frogel *et al.* 1990). The bulge sequence converges with the solar-neighborhood sequence for the bluer stars (down to $(J-H)_0 \approx 0.4$), which reflects the decreasing sensitivity of the infrared colors to metallicity for K giants: in this region of the color-color diagram, globular cluster giants with a range of nearly a factor of 100 in mean metallicity show no discernible trend with [Fe/H] (Frogel *et al.* 1983), but all lie somewhat above and to the left of the mean field line.

The increased scatter of the points with the bluest colors on Figure 12 is consistent with the increasing photometric error at faint magnitudes. We show as an error bar in Figure 12 the mean photometric errors in color for stars with $(J-H)_0 < 0.4$. The systematic shift to the red in $(H-K)_0$ is likely due to an over estimation of the $K$ brightness of stars near our faint limit, §3.2.

Figure 13 is a $(J-K)_0$, $(V-K)_0$ diagram of BW4, where the symbols and solar-neighborhood sequences are as in Figure 12. The upper panel of Figure 13 displays the two colors over the full range present in the data, while the lower panel shows the region of the sequence dominated by the array data. Typical photometric errors for the array data are displayed along the bottom of the lower panel. As was noted by FW87, the M giants lie below the local field giants on this diagram. With our new photometry, we find that the departure from the local sequence begins at $(V-K)_0 \approx 3.75$ which is just at the point where the M stars begin, i.e., just at the point where the TiO bands begin to be strong in the optical and near-infrared portions of the spectrum. The extension of the data blueward of the solar neighborhood giant sequence is likely due to foreground stars (dwarfs and K giants) that have been over corrected for reddening.

Figure 14 shows a color-color diagram in $(V-I)_0$, $(V-K)_0$, where the symbols and principal sequences are again as in Figure 12. Also, as in Figure 13, the upper panel displays the two colors over the full data range while the lower panel shows the region of the sequence dominated by the array data with typical photometric errors in the array data displayed along the bottom. Once again the M giants deviate from the solar neighborhood giant sequence while the bluer K giants do not. It is worth noting that the reddening vectors, shown in Figures 13a and 14a, are nearly parallel to the solar neighborhood giant sequences. Thus errors in reddening cannot cause the deviation found among the M giants.

In previous papers, (FW87, Frogel *et al.* 1990), we argued from two-color diagrams that the displacement of the bulge sequence from the solar-neighborhood sequence was the result of a high mean metallicity among the M giants. This was consistent with the observation that the M giants in Baade's Window have greater absorption in TiO and CO, and in the $K$ band stronger Ca and Na lines than do local field giants. These strong lines were also interpreted to suggest $\langle [Fe/H] \rangle \sim +0.2$ in Baade's Window (FW87; Terndrup *et al.* 1990; Sharples *et al.* 1990; Terndrup *et al.* 1991). In light of our determination of a lower mean metallicity from the slope of the giant branch ($\langle [Fe/H] \rangle \approx -0.28 \pm 0.16$), the strong lines and bands could be interpreted as evidence for selective enrichment in bulge giants (McWilliam & Rich 1994; Terndrup *et al.* 1995). However, as is discussed in §5.2.1, since essentially all bulge giants seem to evolve into M giants, all bulge stars independent of their metallicity) would have to be selectively enriched relative to solar neighborhood stars.

## 5. Bulge Luminosity Functions

### 5.1. Empirical Luminosity Functions

The $K$ magnitude luminosity function (LF) for Baade's Window, derived from the calibrated and dereddened $K$ photometry, is shown in Figure 15 (solid line histogram) and is compared with those derived by DePoy *et al.* (1993), (dashed line histogram) and Davidge (1991), (points). As discussed in §3, our luminosity function is effectively complete for $9.0 \leq K_0 \leq 16$. The upper limit at $K_0 = 9$ is due to image saturation on our short exposure frames. The LF becomes rapidly incomplete for $K_0 > 16$, thus no attempt was made to correct the LF fainter than this limit. To guarantee that there is no magnitude biasing due to different exposure times or seeing conditions on different nights, only the data for BW4b from 1993 July 7, are used to construct the LF.

The DePoy *et al.* (1993) data, which come from an area of 604 arcmin$^2$, have been binned into 0.5 magnitude bins and scaled by 234 to match the area we covered in BW4b, 2.58 arcmin$^2$. The Davidge (1991) data were taken in an area of 1.61 arcmin$^2$, so each point in his LF was multiplied by a factor of 1.61. With just these area corrections, the agreement between the three studies is good. Because of their small areas, both our data and the Davidge (1991) sample suffer from small number statistics in the bins with $K_0 \lesssim 12$. Nevertheless, they both agree, within the errors, even in this magnitude range. Davidge's data begin to differ systematically from ours for $K > 14.5$. Davidge's data are only complete to $K = 13.5$; he applied a completeness correction to points with $K > 13.5$ (open points in Figure 15). Therefore the disagreement between his data and ours is likely due to uncertainties in this correction. The line drawn in Figure 15 is the power law derived from the DePoy *et al.* (1993) data and used in the artificial star experiments. It is a linear fit to their data in the magnitude range $8.9 \leq K_0 \leq 12.1$, and has a slope of 0.278. In general this line matches our data well, especially in the range $14 \leq K_0 \leq 16$, where uncertainty in the number of stars in each bin is smallest.

Figure 16 is a the luminosity function constructed from of the single-channel photometry from FW87 (dotted line histogram), and the combined photometry from DePoy *et al.* (1993) in the range $9 \leq K_0 \leq 12.5$, and our array data in the range $12.5 \leq K_0 \leq 17$. The DePoy *et al.* data are used in the brighter magnitude range because in this range they are a much better sampled set than ours and, as we have just shown, both sets are in good agreement in the region of overlap. The single-channel photometry from FW87 is used in order to extend the luminosity function up to $K_0 \approx 5.5$ since both the DePoy *et al.*

photometry and our photometry are complete up to only $K_0 \approx 9$.

This combined luminosity function, complete over a range of 11 magnitudes in $K$, is the most extensive infrared LF ever published for the bulge. It is well matched over most of this range by a simple power law with a few exceptions. Most apparent is the excess of stars in the range $12.5 \leq K_0 \leq 14$. This excess is due to horizontal branch or clump stars as well as first ascent bump stars. Estimating the number of clump versus bump stars as well as using them to estimate the helium abundance in bulge stars is examined in detail in §5.2.2 below. A second possible feature in the combined luminosity function is the small decrease in the number of stars relative to the power law brighter than $K_0 \approx 8.5$. Although this small decrease could just be due to the small number of bright stars measured in FW87, we note that for a distance modulus of $(m - M)_0 = 14.5$ ($R_0 = 8$ kpc) and the bolometric corrections we derive in §6, the Revised Yale Isochrones (RYI; Green *et al.* 1987) predict the first ascent giant branch tip to be at $K_0 \approx 8.25$, weakly dependent on the choice of age and metallicity. Also, the slope of the AGB LF is expected to be flat (above the tip of the GB) consistent with our empirical LF in the bins $7.0 \leq K_0 \leq 8.5$. If this interpretation is correct, then stars seen at $K_0 < 7$ are likely foreground stars.

In Figure 17, the luminosity function for BW6 is constructed from two sources; the single channel photometry of Frogel *et al.* (1990), (dotted line histogram), and our new array photometry (solid line histogram). The single-channel photometry has been normalized so that the 9.25 magnitude bin matches the derived power law. The straight line is the power law derived by performing a least squares fit to the array data in the range $10 \leq K_0 \leq 16.5$. The derived slope is $0.267 \pm 0.028$, statistically identical to the slope found for BW4. All of the bins brighter than $K_0 \approx 15$ are limited by small number statistics, yet a small excess of stars near $K_0 \approx 12.5$ is suggestive of the HB, clump, and bump stars seen clearly in the BW4 luminosity function. In addition to the smaller number of stars in the sample for BW6, features in the luminosity function are expected to be less distinct than in the BW4 luminosity function because of the increased magnitude dispersion caused by viewing the bulge along a line of sight at greater latitude (Frogel *et al.* 1990).

## 5.2. Comparison with Theory

We now compare the LF derived for BW4 with theory. We will first consider the question of whether the full range of bulge metallicity is represented by the brightest stars in the bulge (e.g., Frogel *et al.* 1990; DePoy *et al.* 1993). We will then attempt to estimate the ratio of HB and clump stars to the number of stars on the first ascent giant branch. This calculation requires that the HB and clump stars be separated from the first ascent bump which falls in the same $K$ magnitude range as the clump for a population like the bulge. With this ratio in hand, we proceed to estimate the helium abundance of the bulge stars with the $R'$ method. The helium abundance influences the masses of turnoff stars at any age and metallicity, especially for metal-rich stars (e.g., VandenBerg & Laskarides 1987; Green *et al.* 1987). The relative enrichment of helium and metals sets constraints on models for the chemical enrichment which operated in the population, and sensitively determines the evolution of the horizontal branch and beyond (e.g., VandenBerg & Laskarides 1987; Greggio & Renzini 1990).

### *5.2.1. Theoretical Luminosity Functions and the Question of Luminous K Stars*

We compare the observed LF to theoretical LFs derived from the Revised Yale Isochrones (Green *et al.* 1987) with the values: $Y = 0.20, 0.30$; $Z = 0.001, 0.004, 0.01, 0.04$; and $t = 5, 10, 15$, where $Y \equiv$ helium abundance, $Z \equiv$ metallicity, and $t \equiv$ age in $10^9$ years. Luminosity functions of representative isochrones are shown in Figure 18. $K_0$ magnitudes were determined from total luminosities along the isochrones using the bolometric $K$ correction determined in §6 and a distance modulus of $(m - M)_0 = 14.5$. A Salpeter IMF was used but the nature of the IMF was found to have little relevance on the giant branch LF. The shape of the observed LF for the bulge (histogram in Figure 18) agrees quite closely with the theoretical luminosity functions for all $Z$ between 0.001 and 0.04 and for all ages between 5 and 15 Gyr. For all of these luminosity functions, the $K_0$ magnitude of the tip of the giant branch lines in the narrow range $7.97 \leq K_0 \leq 8.47$. The tip brightness rises slightly with increasing metal abundance, but varies little with a helium abundance between $0.2 \leq Y \leq 0.3$.

This small variation in the brightness of the GB tip leads to an interesting conclusion. Any brightness difference between the brightest metal poor (K giant) stars and the brightest metal rich (M giant) stars can only be around $\sim 0.5$ magnitudes in $K$. DePoy *et al.* (1993) estimate that all but $\sim 10\%$ of the stars in their Baade's Window survey with $K_0 < 10$ could be identified with known M giants identified in the grism surveys of Blanco *et al.* (1984) and Blanco (1986). The combination of these two facts suggests that not more than $\sim 10\%$ of the stars are metal poor enough to remain K giants up to the tip of the GB. This result agrees with the findings of McWilliam & Rich (1994). They recalibrated the metallicity scale of Rich (1988); their Figure 17 gives a histogram of the distribution of [Fe/H] for Baade's Window. Assuming that a metallicity comparable to that of 47 Tuc ([Fe/H] $\approx -0.75$) is the minimum required for a giant star to become an M giant, then of the 88 stars in their survey 11, or $\sim 12\%$, have [Fe/H] $\leq -0.75$. This is in good agreement with the 10% from DePoy *et al.* and supports the conclusion that while there may be bright K giants in Baade's Window which are part of the bulge population, they constitute not more than 10% of the bright giant population. minority.

### 5.2.2. *First Ascent Bump and Clump Stars and the Bulge Helium Abundance*

Determination of the relative number of bump stars was made using theoretical luminosity functions which were derived from the Revised Yale Isochrones (Green *et al.* 1987) as detailed above in §5.2.1 and are shown in Figure 18. As can be seen from Figure 18, the theoretical giant branch luminosity function is approximately a power law. Including an AGB correction as detailed below, the theoretical power law agrees with the empirical power law within the errors. To calculate the relative number of bump stars, the empirical power law was interpolated under the bump and the excess number of stars was calculated and compared to the excess number of stars found in the observational data. For a population with $Y \sim 0.3$ and $[Fe/H] \sim -0.3$ the first ascent bump was found to contribute $\approx 18\%$ of the observed excess.

In the recent literature, the $R$−method (Iben 1968; Buzzoni *et al.* 1983; Iben & Renzini 1984) or variations thereon have been used to estimate the helium abundance in bulge stars (e.g. Minniti 1995). We summarize these estimates in Table 9. As originally formulated, the R method estimates $Y$ in an old, metal-poor population, for example globular clusters, by the ratio $R \equiv n_{HB}/n_{RGB}$, where $n_{HB}$ and $n_{RGB}$ are the number of stars on the horizontal branch and on the red giant branch, respectively, which have luminosity greater than the horizontal branch RR Lyrae stars. This ratio is assumed to equal the ratio of the time spent by stars on the horizontal and red giant branches. The latter ratio is calibrated using stellar evolution theory, leading to $Y = 0.380 \log R + 0.176$ (Buzzoni *et al.* 1983).

Although the bulge is old enough for the $R$−method to be valid (Renzini 1994), the mixing of several evolutionary stages in both color and magnitude space on an IR CMD make the application of this method to the bulge rather difficult. First, the metallicity of the bulge is high enough that most of the horizontal branch stars are in a clump near the giant branch, and consequently the second-ascent stars will have colors which are nearly the same as first-ascent stars. Second, due to the front-to-back distance dispersion and the metallicity dispersion of the stars in bulge fields, it is impossible to completely separate first ascent giants from second ascent asymptotic giant branch stars. Finally, as described above, the horizontal branch is mixed in with the first-ascent bump. Therefore we are not able, as would be the case in globular clusters, to count stars on the horizontal, giant, and asymptotic branches to determine the value of $R$. Instead we need to make reasonable guesses about the fraction of stars in different stages of evolution.

A partial solution to this problem is given by the $R'$−method, a variation on the $R$−method which determines the ratio

$$R' \equiv \frac{n_{\rm HB}}{n_{\rm RGB} + n_{\rm AGB}},$$

where $n_{\rm AGB}$ is the number of stars on the asymptotic giant branch. The calibration of $Y$ with $R'$ is $Y = 0.457 \log R' + 0.204$ (Buzzoni et al. 1983). We adopt this method here.

We derived the value of $Y$ for the Baade's Window fields as follows: Since the luminosity function is approximated well by a power law (§5.1, above), we fit a power law to the luminosity function excluding the three bins containing the HB, (the HB is contained in the three bins spanning the range $12.5 < K_0 < 14.0$), and subtracted the fit to derive $n_{\rm HB}$. Before actually fitting the power law, we assumed that 25% of the stars in each bin above the HB were on the asymptotic giant branch (Terndrup 1988), and fit the power law after subtracting 25% from each bin. We count GB stars ($n_{GB}$) down to the top of the clump. We derive $R' = 1.58$ and $Y = 0.30 \pm 0.03$. The error assigned to $Y$ is from counting statistics in the luminosity function. For $Y \sim 0.3$, the luminosity function derived from the Revised Yale Isochrones indicates that about 18% of the stars in the region of the horizontal branch clump are likely to be stars in the first-ascent bump (§5.3.2). This reduces $R'$ to 1.40 from which we derive $Y = 0.27 \pm 0.03$.

We can compute the ratio $\Delta Y/\Delta Z$ for the bulge, assuming the primordial helium abundance is $Y_p = 0.235$ (Pagel et al. 1992; Yang et al. 1994) and $Z_\odot = 0.02$ (Green et al. 1987). For the mean abundance in the bulge, we adopt $\langle Z \rangle \sim 0.52 Z_\odot$, which is derived from our mean bulge abundance of $\langle[\text{Fe/H}]\rangle = -0.28$, above. This gives $\Delta Y/\Delta Z = 3.3 \pm 1.3$, in line with recent estimates from optical CMDs (Renzini 1994). Without the correction for first-ascent bump stars, we would have derived $\Delta Y/\Delta Z = 6.2 \pm 1.5$.

While the method we have just detailed gives results that agree with others from the literature (Table 9), we note that one of our parameters, the lower magnitude limit on $n_{GB}$, is theoretically ill defined for a metal rich population. In both the $R$ and $R'$ methods, the lower magnitude limit for $n_{GB}$ is defined as the magnitude of the RR Lyrae stars and the methods are calibrated with metal poor populations. Since the bulge RR Lyrae are representative of only the metal poor tail of the metallicity distribution in the bulge (Walker & Terndrup 1991), this limit cannot be used for the bulge population as a whole. In actuality the magnitude of RR Lyrae stars is a weak function of metallicity. Lee et al. (1990) derived the equation $M_{Bol}^{RR} = 0.20[\text{Fe/H}] + 0.81$. However, if we use this on our sample (giving $K_0^{RR} = 13.4$) we derive $Y = 0.18$, a clearly unphysical value. Until the $R$ and $R'$ methods are calibrated for metal rich populations, the validity of applying them to the bulge is unclear.

## 6. The Tonry-Schneider Distance Indicator

A distance indicator of potentially high accuracy for galaxies, which measures the spatial surface brightness fluctuations in early-type galaxies, was introduced by Tonry & Schneider (1988). If the stellar luminosity functions of early-type galaxies (or the bulges of spirals such as M31) are similar, then relative distances can be measured to high accuracy (5%) since the images of distant galaxies would be smoother than those of nearby systems.

The surface brightness fluctuation method (SBF) is thoroughly described by Ajhar & Tonry (1994). In brief, Tonry and Schneider introduced the mean, luminosity-weighted luminosity of a stellar population, defined as

$$\bar{L} = \frac{\sum n_i L_i^2}{\langle L \rangle},$$

where

$$\langle L \rangle = \sum n_i L_i.$$

Here $n_i$ is the number of stars in the $i$th species, each of which has luminosity $L_i$ in a particular bandpass. In the SBF technique, the quantity $\bar{L}$ is converted to a magnitude $\bar{m}$, and compared to a theoretical value $\bar{M}$ or one determined empirically through observations of nearby galaxies. The distance $d$ in parsecs is then derived through

$$\bar{m} = \bar{M} + 5\log_{10}\left(\frac{d}{10\mathrm{pc}}\right).$$

The success of the SBF method depends critically on having a reliable $\bar{M}$. The original calibration adopted by Tonry and Schneider used isochrones from the RYI to compute $\bar{M}$. Later, Tonry (1991) adopted an empirical calibration, using observations of the fluctuation statistics of M32 and the bulge of M31 to derive

$$\bar{M}_I = -4.84 + 3.0(V - I)_0,$$

where $(V - I)_0$ is the dereddened integrated $V - I$ color of the stellar population. This calibration has a dependence on color that is stronger and of the opposite sign from the RYI calibration. The source of this difference was discussed by Tonry (1991), who speculated that the bolometric corrections used in the RYI were too small. This was confirmed by Ajhar & Tonry (1994), who noted that the giant branches of the RYI fail to turn over at high metallicity, and who argued that the behavior of $\bar{M}_I$ with color in metal-rich globular clusters was consistent with the empirical M31/M32 calibration.

How does the bulge fit into all this? With our new luminosity function we can test whether the stellar content of the bulge is similar to that of the bulge of M31 and of the central regions of other galaxies. Our new infrared photometry allows us to compute bolometric corrections for K and M giants in the bulge and to compare these to the values in the RYI. We also have a good determination of the luminosity function in the bulge, which allows us to calculate $\bar{M}_I$ and see how this compares to that found for M31.

We proceed by deriving bolometric magnitudes for the stars in the BW4a and BW4b fields using the method described by FW87, adopting the sign convention

$$M_{\rm bol} = M_V + BC_V$$
$$= M_I + BC_I.$$

Figure 19 displays the bolometric corrections in the $V$ and $I$ bands as a function of $(V-I)_0$. The open points show the data from this paper, while the $\times$ symbols show the data from FW87.

For the M giants in Figure 19, we simply plotted the bolometric corrections from FW87, which were computed using optical photographic photometry and infrared single-channel data. We can check the accuracy of this photometry by comparing the $V$ and $I$ magnitudes from CCD surveys to the values adopted by FW87. The $V$ data were from Blanco (1986), who applied a correction given by Blanco & Blanco (1986) to Arp's (1965) photographic photometry. The $I$ photometry in FW87 was from photographic photometry supplied by Whitford. The differences between the $V$ and $I$ magnitudes in FW87 and those from the more accurate CCD surveys is not significant:

$$\langle I({\rm CCD}) - I({\rm FW87}) \rangle = -0.16 \pm 0.11 ({\rm s.d.}),$$
$$\langle V({\rm CCD}) - V({\rm FW87}) \rangle = +0.09 \pm 0.33 ({\rm s.d.}).$$

Because these differences are small, we decided not to recompute $M_{\rm bol}$ for the M giants in

- 29 -FW87 using the new CCD photometry. The effect would have been to increase the values of $M_{bol}$ by only 0.04, mainly due to the brighter $I$-band scale of the CCD photometry.

In Figure 19, we also compare these empirical bolometric corrections to the RYI values (thin solid line). We also have plotted (thick solid line) the empirical bolometric corrections for globular cluster stars (Da Costa & Armandroff 1990) and (dashed line) empirical values for local stars presented by Bessell & Wood (1984). The bolometric corrections for the bulge stars are in excellent agreement with both empirical derivations for the warmer K stars ($V - K \leq 2$), but have values, up to 0.4 magnitudes greater for cooler stars. This is most likely a consequence of the stronger molecular absorption in bulge giants than in local stars of the same temperature (Terndrup *et al.* 1990), which suppresses the $V$ and $I$ flux in the former stars. The bolometric corrections adopted in the RYI are clearly incorrect, with values that are much too small for stars with $V - I > 2$. This difference between the bulge and RYI bolometric corrections is just what is expected from the empirical calibration of the Tonry-Schneider method (Tonry 1991). That the bulge bolometric corrections are large implies that the bulge stars are fainter in the $V$ and $I$ bands than the RYI isochrones, as seen in metal-rich populations (e.g. Ajhar & Tonry 1994).

As a test that the LFs for the Galaxy's and M31's bulge are similar, particularly for RGB and AGB stars, we now estimate the value of $\bar{m}_I$ for the Galaxy's bulge. This value is found by separately calculating $\sum n_i L_i^2$ and $\sum n_i L_i$. The first term is sensitive to the steepness of the luminosity function at the very top of the giant branch, while the second term is the total light of *all* stars. A similar calculation gives $\bar{m}_V$.

To compute these quantities, we obtained $V$- and $I$-band luminosity functions (cf. Terndrup *et al.* 1990) derived from deep CCD surveys in several bulge fields, that were already corrected for incompleteness and foreground contamination (Terndrup 1988). We also eliminated stars with $I < 11.5$, which are almost certainly foreground giants (Walker

& Mack 1986; Terndrup *et al.* 1995). We derive a mean, extinction-corrected color for the Baade's Window field of $(V-I)_0 = 1.09 \pm 0.10$, slightly bluer than the value of 1.18 for M31's bulge (Ajhar & Tonry 1994). We also find $\bar{m}_I = 12.85 \pm 0.15$, again corrected for reddening. The difference between the fluctuation magnitudes in $V$ and $I$ for the bulge is $(\bar{V}-\bar{I})_0 = 1.96 \pm 0.10$, again slightly bluer than the M 31 value of $2.26 \pm 0.14$.

From the calibration of $\bar{M}_I$ with integrated color (Tonry 1991), we can derive a distance modulus of $14.43 \pm 0.35$ to the bulge ($R_0 = 7.7 \pm 1.2$ kpc), which is very similar to most other recent distance estimates to the galactic center (Reid 1993). The relatively large error in the distances arises partly from the uncertainty in the reddening, but mostly from the strong dependence of $\bar{M}_I$ on color. Since the reddening towards the bulge is large and spatially variable (Blanco *et al.* 1984), it is unlikely that the error in the reddening and therefore in the integrated $(V-I)_0$ color will ever be reduced to a sufficiently small value that the bulge can be used as a *primary* calibrator of the distance fluctuation method. Nevertheless, we have demonstrated that the bulge's luminosity function and the relative faintness of its M giants in $V$ and $I$ (compared to the RYI) are two characteristics which it shares with most other galaxies.

## 7. Summary and Conclusions

In this paper we present deep $JHK$ photometry of Baade's Window (BW4) and of a field on the minor axis at b = $-6°$ (BW6). In order to estimate the completeness of our data as a function of magnitude and to determine the faint limit, we constructed entirely artificial frames, paying careful attention to modeling the level of crowding, unresolved background, sky level, and noise. Photometry of stars in these artificial frames was then measured with DoPHOT in a manner identical to that used for the real frames. We found that the photometry was complete to $K_0 = 16$ and that the average photometric accuracy to this limit was better than 0.04 magnitudes. In addition we were able to determine that the errors reported by DoPHOT are an accurate representation of the true uncertainties in the measurements over the entire range sampled.

We constructed $(J - K)_0$, $K_0$ color-magnitude diagrams of the BW4a, BW4b and BW6 fields. The CMDs for both the BW4b and BW6 field are deeper ($K_0 \approx 16.7$, complete to $K_0 = 16$) than previously published diagrams. By combining our array data with the single-channel photometry from Frogel *et al.* (1984), FW87, and Frogel *et al.* (1990), we were able to extend the CMDs up to $K_0 \approx 5.5$. Our total coverage is then, $5.5 \leq K_0 \leq 16.7$, a range larger than in any previous study. By combining the IR photometry of the BW4b field with optical data from Terndrup & Walker (1994), we made a comparison of optical and infrared CMD morphological features. We find a blue foreground dwarf sequence in the IR (from $K_0 \sim 13$ to $K_0 \sim 16.5$ with $(J - K)_0 \leq 0.45$) analogous to that seen in the optical, in contrast to claims of its absence by Davidge (1991). We also find a GB clump in the IR ($12.5 \lesssim K_0 \lesssim 14.0$) in which all of the stars with optical data fall in the optical GB clump. This clump is particularly noticeable in the LF and is discussed below.

We used the relation between the slope of the upper giant branch and [Fe/H], found by Kuchinski *et al.* (1995) for globular clusters, to calculate $\langle$[Fe/H]$\rangle$ for BW4a, BW4b, and BW6. For the BW4 fields we derived $\langle$[Fe/H]$\rangle = -0.28 \pm 0.16$, a value close to that found by McWilliam & Rich (1994) from a spectroscopic study of K giants, but less than estimates for $\langle$[Fe/H]$\rangle$ from other metallicity sensitive near-IR parameters and from TiO absorption bands. McWilliam & Rich have suggested that selective elemental enhancement could be responsible for the strong TiO bands. They also point out several possible explanations for the enhanced CO absorption seen by FW87. From the discussion in Frogel (1993) however, we note that interpretation of near-IR colors and band strengths in terms of [Fe/H] may be more ambiguous than previously thought.

To extend the Kuchinski *et al.* (1995) technique of measuring average [Fe/H] to various latitudes in the bulge, we also applied it to the fields from Frogel *et al.* (1990). Our results are summarized in Tables 7 and 8. The metallicity gradient along the minor axis found from these fields is $-0.060 \pm 0.033$ dex/degree ($-0.43 \pm 0.21$ dex/kpc for $R_0 = 8$ kpc). As shown in Figure 11 and compiled in Table 8, this value of the metallicity gradient agrees reasonably well with values from the studies of Terndrup (1988), Frogel *et al.* (1990), and Tyson (1991). This agreement between the studies can also be seen on a field by field basis, which suggests that the giant branch slope-[Fe/H] relation, which was derived for metal rich globular clusters, is also applicable to the Bulge.

We examined $(J-H)_0$, $(H-K)_0$; $(J-K)_0$, $(V-K)_0$; and $(V-I)_0$, $(V-K)_0$ diagrams for the BW4b field. All showed the same behavior. Consistent with previous studies (FW87; Frogel *et al.* 1990) the bulge M giants are found to lie below the solar neighborhood giant relations in these diagrams. However, since our data go much deeper and hence much bluer than previous work, we were able to determine that this departure from the mean colors for solar neighborhood giant behavior is not true for K giants. The bulge K giants follow the solar neighborhood giant sequence. This convergence may be due to the decreasing sensitivity of the infrared colors to metallicity (at least for the IR color-color diagram), or it may be that selective enrichment of Ti found in Bulge stars (McWilliam & Rich 1994; Terndrup *et al.* 1995) causes the TiO bands, found in M giants and not in K giants, to significantly affect the colors of Bulge M giants.

Combining our array data with the single-channel photometry from Frogel *et al.* (1984), FW87, and Frogel *et al.* (1990) and the array data from DePoy *et al.* (1993), we construct LFs for the BW4b and BW6 fields. The BW4b LF is complete over the range $5.5 \leq K_0 \leq 16.0$, and the BW6 LF is complete over the range $6.5 \leq K_0 \leq 16.5$. Both LFs generally follow a power law, except for excesses of stars in the range $12.5 \leq K_0 \leq 14.0$ and at magnitudes greater than the giant branch tip ($K_0 \approx 8.25$). These IR LFs are the most extensive currently published for the bulge.

The empirical LFs agree closely with theoretical (RYI) LFs of comparable metallicity, age, and helium abundance. The first ascent giant branch tip luminosity is found to vary only about 0.5 magnitudes, between $7.97 \leq K_0 \leq 8.47$, for reasonable variations in age and overall metallicity. This fact suggests that both stars metal poor enough to remain K giants to the tip of the giant branch and stars metal rich enough to become M giants must not differ by more than $\sim 0.5$ magnitudes in $K$ at the tip of the giant branch. DePoy *et al.* (1993) find that all except $\sim 10\%$ of the bright stars observed in Baade's Window correspond with M giants detected on the grism surveys of Blanco *et al.* (1984) and Blanco (1986). This 10% agrees well with the percentage of bulge stars expected to have metallicities low enough to remain K giants to the tip of the GB (Rich 1988, McWilliam & Rich 1994).

We derive a helium abundance for Bulge stars of $Y = 0.27 \pm 0.03$ with the $R'$ method. From our mean metallicity, $\langle [\text{Fe/H}] \rangle = -0.28$, we adopt $\langle Z \rangle = 0.52 Z_\odot$ which gives $\Delta Y / \Delta Z = 3.3 \pm 1.3$, a value in agreement with other recent estimates. In spite of this agreement, we point out that the lower giant branch limit from which $n_{GB}$ is calculated in both the $R$ and $R'$ methods is ill defined for a metal rich population. The magnitude of the RR Lyrae stars cannot be used in the bulge because the RR Lyraes represent only the metal poor tail of the metallicity distribution. Until the $R$ and $R'$ methods are calibrated for metal rich populations, the validity of applying them to the Bulge is unclear.

We used our optical and IR photometry to calculate bolometric magnitudes for bulge K and M giants using the method detailed in FW87. The resulting bolometric corrections are up to 0.4 magnitudes greater than those used in the RYI for cool M giants. These higher values agree with the empirical findings for globular cluster stars (Da Costa & Armandroff 1990) and local disk stars (Bessell & Wood 1984). We then use the bolometric magnitudes and the bulge LF to calculate a distance to the bulge using the Tonry–Schneider surface brightness fluctuation method. We derive a value of $(m-M)_0 = 14.43 \pm 0.35$ ($R_0 = 7.7 \pm 1.2$ kpc). This value of $R_0$ is very similar to most other recent distance estimates to the galactic center.

JAF thanks Leonard Searle for the research opportunities provided by a Visiting Research Associate at LCO. JAF's bulge research is supported by NSF grant AST92-18281. DMT's research is supported in part by NSF grant AST91-57038. We are particularly grateful to S. E. Persson for the use of IRCAM (built with NSF funds) at LCO and also thank Miguel Roth, Bill Kunkel, and members of the LCO support crew for their able assistance in the set up and operation of IRCAM and the duPont and Swope telescopes. We appreciate the advice given by Paul Schechter on the use of DoPHOT which was developed with the aid of NSF grant AST83-18504. Our colleagues at OSU, Andy Gould, Mark Houdashelt, Kris Sellgren, and David Weinberg made a number of helpful suggestions that resulted in a significantly better presentation.

Fig. 1.— Representative $K$ band exposures of our fields. Coordinates are in pixels with arbitrary zero points. The labeled stars are from Blanco *et al.* (1984) and were used to calibrate the frames using the photometry of FW87. (a) An image of BW4a. (b) A deep image of BW4b. The artifacts circled in white are electronic "ghosts" caused by saturated pixels in the NICMOS 3 chip. Stars falling in such regions were removed after photometric processing. (c) A deep image of BW6. (d) An artificial image designed to mimic the BW4b field.

Fig. 2.— The mean difference between the measured and input magnitude as a function of input magnitude for the artificial data frame. Error bars are a measure of the scatter in each 0.25 magnitude bin. The saturation limit for the real data is indicated.

Fig. 3.— Errors as a function of magnitude. (a) The absolute value of the True Errors vs magnitude. (b) The errors assigned by DoPHOT vs magnitude. The histogram is the average of the true errors binned in 0.25 magnitude bins.

Fig. 4.— (a) Input luminosity function (solid line histogram) and detected luminosity function (dotted line histogram). Scatter in the input luminosity function is due to counting statistics in each bin. (b) Completeness as a function of magnitude. Few bins are 100% complete due to the way DoPHOT handles saturated stars. Therefore the 93% level is considered complete (see text.)

Fig. 5.— A comparison of the real (solid line histogram) and simulated (dotted line histogram) luminosity functions. Neither histogram has been normalized. Note that to within counting statistical error the histograms agree down to $K \approx 16.5$.

Fig. 6.— $(J - K)_0$, $K_0$ color-magnitude diagrams of BW4a and BW4b. Solid circles are photometry from this paper. Open squares are single channel photometry from FW87. Pluses are photometry from Frogel *et al.* (1984). The lines are least squares fits to the upper giant branch in the range $8.0 \leq K_0 \leq 12.6$.

Fig. 7.— $(J - K)_0$, $K_0$ color-magnitude diagram of BW6. Solid circles are photometry from this paper. Open squares are from Frogel *et al.* (1990). The line is as in Figure 6.

Fig. 8.— $(V - I)_0$, $I_0$ optical color-magnitude diagram with selected regions indicated (data from Terndrup & Walker 1994). Region I is the foreground dwarf sequence, and Region II is the clump.

Fig. 9.— $(J - K)_0$, $K_0$ IR color magnitude diagram with stars corresponding to the optical regions from Figure 8 highlighted. The solid squares correspond to Region I (foreground sequence) stars and the solid circles correspond to Region II (clump) stars.

Fig. 10.— $(J - K)_0$, $K_0$ color magnitude diagrams of the minor axis fields from Frogel *et al.* (1990). The lines are least square fits to the giant branches in the range $8.0 \leq K_0 \leq 12.6$.

Fig. 11.— Metallicity gradient in the Bulge as found in four independent studies. In all panels the solid line is the least squares fit to the indicated data, and the dotted line is the least squares fit to all of the data from the five studies combined. (a) Metallicity values from this work, from fits to the giant-branch slopes of Frogel *et al.* (1990) data (open circles) and the metallicity value from McWilliam & Rich (1994) (asterisk, included for comparison with our [Fe/H] value for BW4). The solid line is a least squares fit to our data only. (b) Metallicity values from Frogel *et al.* (1990) based on CO strengths. (c) Metallicity values from Terndrup (1988). (d) Metallicity values from Tyson (1991).

Fig. 12.— $(J-H)_0$, $(H-K)_0$ diagram of stars from BW4. The solid circles are known dwarfs from the foreground sequence. The ×'s are single-channel photometry from Frogel *et al.* (1984) and FW87. The lines are sequences for solar neighborhood disk stars from Bessell & Brett (1988). The solid line is for giants and the dotted line is for dwarfs. Typical errors for stars at the blue end of the sequence are indicated at the lower right. The vector indicates the applied reddening correction.

Fig. 13.— $(J-K)_0$, $(V-K)_0$ diagram of stars from BW4. Symbols and lines are as in Figure 12. (a) The entire data range is shown. The vector indicates the applied reddening. (b) The data range of the array data alone is shown. Typical photometric errors are displayed along the bottom of the diagram.

Fig. 14.— $(V-I)_0$, $(V-K)_0$ diagram of stars from BW4. Symbols and lines are as in Figure 12. (a) The entire data range is shown. The vector indicates the applied reddening. (b) The data range of the array data alone is shown. Typical photometric errors are displayed along the bottom of the diagram.

Fig. 15.— Comparison of the BW4b (solid line histogram), DePoy *et al.* (1993) (dotted line histogram), and the Davidge (1991) (points) luminosity functions after normalizing to an area of 2.58 arcmin$^2$. The open points in the Davidge LF have been corrected, by Davidge, for incompleteness. The line is a power law fit to the DePoy *et al.* data.

Fig. 16.— Comparison of the FW87 luminosity function (dotted line histogram) and the combined luminosity function of the DePoy *et al.* (1991) and our data (solid line histogram). The line is a power law fit to the DePoy data.

Fig. 17.— The luminosity function of BW6. The dotted line histogram is for data from Frogel *et al.* 1990 and has been scaled so that the 9.25 magnitude bin matches the derived power law. The dotted line histogram is from our array photometry. The line is the power law derived from a least squares fit to the array data in the range $10 \leq K_0 \leq 16.5$.

Fig. 18.— The giant branch luminosity functions from the Revised Yale Isochrones (Green *et al.* 1987) for stars with bulge parameters. Note the first ascent bumps near $K_0 \approx 13$ as well as the fact that the magnitude of the GB tip is not a strong function of metallicity. $Z \equiv$ Metallicity, $Y \equiv$ Helium Abundance, $t \equiv$ Age in Gyr. The histogram is the empirical combined data from Figure 16. The normalization of this histogram is arbitrary, but note the excellent agreement in slope.

Fig. 19.— $V$ and $I$ bolometric corrections as a function of $(V-I)_0$. The open points show the data from this paper, while the ×'s show data from FW87. The thin solid line is the corrections used in the RYI. The thick solid line is the empirical bolometric corrections for globular clusters stars from Da Costa & Armandroff (1990). The dashed line is the empirical bolometric corrections for local stars from Bessell & Wood (1984).

– 43 –